\documentclass[prl,twocolumn,amsmath,amssymb]{revtex4}
\usepackage{graphicx}
\begin{document}

\title{Aging under Shear: Structural Relaxation of a Non-Newtonian Fluid}
\author{R.~Di Leonardo}
\email{roberto.dileonardo@phys.uniroma1.it}
\author{F.~Ianni}
\author{G.~Ruocco}
\affiliation{Universit\'a di Roma ``La Sapienza'' and INFM,
I-00185, Roma, Italy.}
\date{\today}

\begin{abstract}

The influence of an applied shear field on the dynamics of an aging colloidal suspension has been
investigated by the dynamic light scattering determination of the density autocorrelation function. 
Though a stationary state is never observed, the slow 
dynamics crosses between two different non-equilibrium regimes as soon as the structural
relaxation time $\tau_s$ approaches the inverse shear rate $\dot{\gamma}^{-1}$.
In the shear dominated regime (at high $\dot\gamma$ values) the  structural relaxation time is found to be strongly
sensitive to shear rate ($\tau_s\sim\dot{\gamma}^{-1}$) while aging proceeds at a very slow rate.
The effect of shear on the detailed shape of the density autocorrelation function is quantitatively described 
assuming that the structural relaxation process arises
from the heterogeneous superposition of many relaxing units each one
independently coupled to shear with a parallel composition rule for
timescales: $1/\tau\rightarrow1/\tau+A\dot\gamma$. 

\end{abstract}

\maketitle

One of the most peculiar and intriguing behaviour of soft materials is the
strong sensitivity of their flow properties to the application of an even
slight external deformation \cite{larson}.
When a liquid flows in a steady shear state, the inverse shear rate
$\dot{\gamma}^{-1}$ introduces a new relevant timescale in the dynamics.  
On a microscopic level a non-Newtonian character reflects a competition
between $\dot{\gamma}^{-1}$ and the natural timescale of those particle
rearrangements controlling  the macroscopic flow properties.
For moderate shear rates, i.e. those occurring in the macroscopic world, mainly the slow
modes are affected by shear and give rise to complex rheological behaviour.
Understanding the physical mechanism governing the interaction between slow
dynamics and shear has been the subject of great theoretical and numerical
efforts in recent years.

On the theoretical side, some of the most powerful tools for the investigation
of slow dynamics in complex systems, such as mode coupling theory
\cite{fuchs,miyazaki}, mean-field models \cite{bbk}, trap models \cite{sollich} and
molecular dynamics simulations \cite{yamamoto, berthier, angelani}, have been
recently extended to account for the presence of shear.  The general picture
emerging from these studies indicates  that the  structural dynamics is very
sensitive to even moderate shear rates whenever $\dot{\gamma}^{-1}$ becomes of
the same order of the characteristic timescale for structural rearrangements.
In particular, even starting from non equilibrium, aging states, the presence
of shear ensures the existence of a stationary state whose correlation
functions decay to zero on a timescale governed by $\dot{\gamma}^{-1}$.

Despite the growing amount of numerical and theoretical work investigating
the shear-influenced slow dynamics, an experimental microscopic counterpart is still
relatively poor.  Evidences for a shear dependent structural relaxation time
have been obtained by  diffusing wave spectroscopy \cite{bonnreju,
viasnoff} and Light Scattering Echo experiments \cite{petekidis}, though it is not
straightforward to deduce the microscopic dynamics from the observables probed in
the multiple scattering regime. 
No attempt has been made up to now to use dynamic light scattering (DLS) in the
single scattering regime, a technique that directly probes the density autocorrelation
function, which plays a central role in both theoretical and numerical approaches.

In this Letter we investigate the evolution of the density autocorrelation function
of an aging colloidal suspension subject
to a steady shear flow.  The sample is an aqueous suspension of Laponite, a
highly thixotropic liquid which  undergoes structural arrest on a timescale which
strongly depends on concentration and ionic strength \cite{mourchid} and that can be as long
as few months \cite{barbara}.  We found that
the aging dynamics displays two different regimes whose boundary is marked by the condition $\tau_s
\dot{\gamma}\sim 1$.  As long as the characteristic relaxation time $\tau_s$ is
small on the time-scale $\dot{\gamma}^{-1}$, aging is unaffected by the presence
of shear. During aging dynamics slows down, and when $\tau_s$ becomes of the order of
$\dot{\gamma}$, the system enters a shear dominated regime where aging is
strongly reduced and the structural relaxation time is very sensitive to
$\dot{\gamma}$.  The intermediate scattering functions, characterising the slow
non-equilibrium dynamics of the sheared sample, are well described assuming an
heterogeneous scenario where the complex dynamics results from the superposition of
relaxing units each one independently coupled to shear rate with a parallel composition
rule for timescales: $1/\tau\rightarrow1/\tau+A\dot\gamma$.

\begin{figure} \includegraphics[width=.40\textwidth]{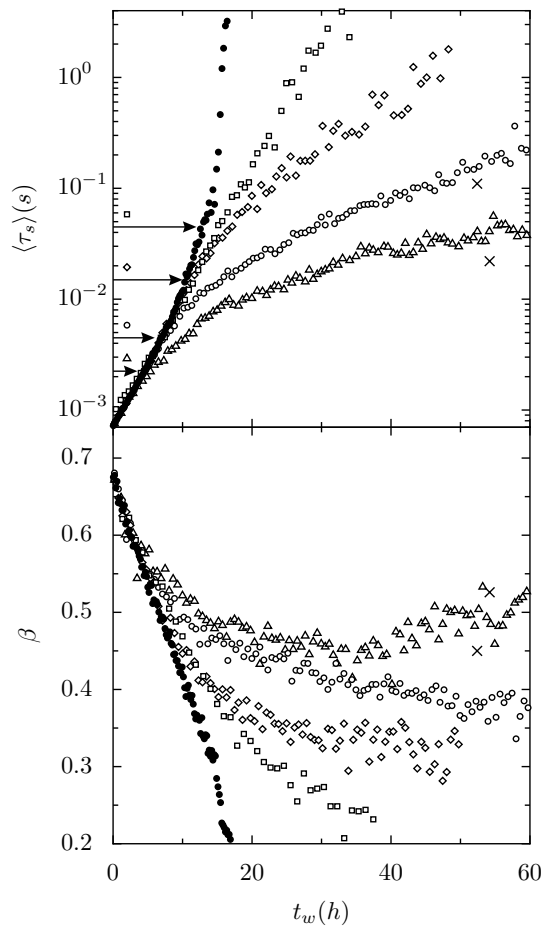} \caption{ Average
slow relaxation time $\langle\tau_s\rangle$ and stretching exponent $\beta$ as 
a function of waiting time $t_w$ during aging under different shear rates $\dot{\gamma}$:
$(\vartriangle)446,\;(\circ)223,\;(\diamond)67,\;(\Box)22\;\textrm{s}^{-1}$.
Solid symbols ($\bullet$) refer to aging without shear. Arrows in top frame indicate
the $\dot{\gamma}^{-1}$ values corresponding to each curve}\label{taubeta}
\end{figure}

The system consists of an aqueous suspension of Laponite RD, a synthetic
layered silicate provided by Laporte Ltd.  Particles are disk shaped with a
diameter of $25$ nm and $1$ nm thickness. Laponite powder is dispersed in
ultrapure water at $3\%$ wt concentration, stirred for $\sim30$ min and then
filtered through a Millipore Millex-AA $0.45$ $\mu$m filter. The obtained
suspension, which is optically transparent and initially liquid, is loaded into
a home made  cone and plate shear cell having a flat optical window as the static
plate. Incident laser beam (He-Ne, $10$ mW) and scattered light pass through
the same optical window. The scattered light is collected by a mono-mode optical fiber,
detected by a photomultiplier  and analyzed by a home
made software correlator, after being optionally mixed with a coherent local
oscillator field. The scattering geometry is fixed (scattering vector $q=25\;\mu\textrm{m}^{-1}$).
The possibility of choosing an heterodyne correlation scheme
enables direct access to the detailed velocity profile in the shear cell. Wall
slip and distortion from linearity in the velocity profile are found to be negligible in the
whole shear rate range here investigated.  The non-equilibrium dynamics of
Laponite suspensions flowing in a ``steady" shear state is then constantly
monitored through the normalized intensity autocorrelation function
$g^{(2)}(t_w, t)=\langle I(q,t_w)I(q,t_w+t) \rangle/\langle I(q, t_w)\rangle^2$.
Where $\langle..\rangle$ indicates temporal average over the acquisition time $T$ which is always
much longer than the characteristic slow relaxation time $\tau$ ($T/\tau\gtrsim 10^2 $)  
but also much shorter than the time
one should wait before changes in $\tau$ are significant ($T\;\partial\log\tau/\partial t_w \lesssim 10^{-2}$).
In the single scattering regime and within the Gaussian approximation
$g^{(2)}(t_w,t)=1+|F_q(t_w,t)|^2$ \cite{berne} where $F_q(t_w,t)=\langle
\rho_{-q}(t_w)\rho_q(t_w+t) \rangle/\langle \rho_{-q}(t_w)\rho_q(t_w)\rangle$ is the
intermediate scattering function of the colloidal particles. 
In order to avoid purely geometric
decorrelation of scattered light, due to both advection and transit time effects
\cite{ackerson}, steady shearing is interrupted every ten minutes and the correlation functions 
are collected in short (few tens of seconds) time intervals during which the rotor
is stopped ($\dot{\gamma}=0$).
The robustness of the results with respect to changes in
acquisition protocol has been checked. 
\begin{figure} \includegraphics[width=.40\textwidth]{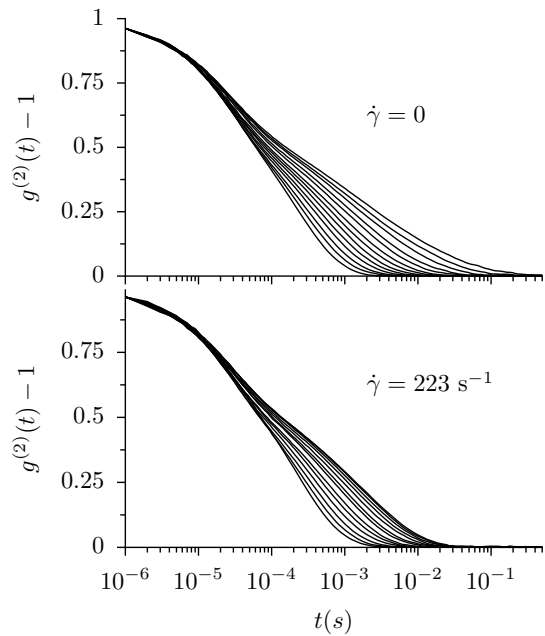} \caption{Normalized
intensity autocorrelation function for ten equally spaced waiting times between
$0.1$ and $15$ hours. Top frame refers to aging without shear while bottom
frame refers to aging under shear with $\dot{\gamma}=223\;
\textrm{s}^{-1}$}\label{corr} \end{figure}
The top frame of Fig. \ref{corr} shows the evolution of $g^{(2)}(t_w,t)$, in the
absence of shear, for an evenly spaced set of waiting times $t_w$ spanning the
interval $0.1-15$ hours.  As already observed in \cite{bonnaging},  aging
dynamics displays a two step relaxation scenario: i) a fast exponential
relaxation process related to single particle diffusion and whose
characteristic time $\tau_f$ remains of the same order of magnitude found in
very dilute suspensions; ii) a slow stretched exponential process related to
cooperative rearranging motions and whose characteristic time $\tau_s$ grows
exponentially fast with waiting time $t_w$ while becoming strongly stretched
(stretching parameter $\beta$ down to  0.2).
In the bottom frame of Fig. \ref{corr} we reported the evolution of
$g^{(2)}(t_w,t)$, observed on the same set of $t_w$ reported in top frame,
when a shear rate $\dot{\gamma}=223$ s$^{-1}$ is applied.  Though a stationary
state is never reached in the observation time window, for long enough waiting
times, the slow relaxation time grows  slower than exponentially (the correlation functions
come closer), while the shape of the relaxation function approaches a constant
profile (constant stretching exponent).  
The two step decay for $F_q(t_w,t)$:
$$
F_q(t_w,t)=f\exp\left[-(t/\tau_s)^\beta\right]+(1-f)\exp\left[-t/\tau_f\right]
$$
where all parameters ($f$, $\tau_s$, $\beta$, $\tau_f$) depend on $t_w$ and on $q$, 
provides a very good fit for all the correlators corresponding to the 
different $\dot{\gamma}$, $t_w$ values here investigated.
The presence of a shear induced
crossover becomes more evident if we plot the fitting parameters $\langle\tau_s\rangle$ 
(average slow relaxation time $\tau_s/\beta\;\Gamma(1/\beta)$, where $\Gamma$ is
the Euler gamma function) and $\beta$  (stretching exponent) as a
function of $t_w$ for different shear rates as in Fig. \ref{taubeta}. 
As long as the system's dynamics is fast on the timescale  $\dot{\gamma}^{-1}$
(indicated by arrows in the plot), non-equilibrium dynamics takes place as if
shear were not present: both the relaxation time $\tau_s$ and the corresponding
stretching exponent $\beta$ evolve with waiting time following the
$\dot{\gamma}=0$ curve (full circles).  The presence of shear starts to  affect the
dynamics as soon as $\tau_s$ becomes larger than $\dot{\gamma}^{-1}$: the growth
of $\tau_s$ is dramatically reduced (even if not completely stopped) and
the stretching parameter $\beta$ behaviour flattens.  Though aging is never
completely absent, the slow relaxation dynamics, for a given waiting time,
appears to be very sensitive to $\dot{\gamma}$ being faster and less
stretched as $\dot{\gamma}$ increases. If we suddenly increase $\dot{\gamma}$,
rejuvenation is observed leading to faster relaxation time and smaller
stretching (higher $\beta$).  Crosses in Fig. \ref{taubeta} are the
average relaxation times and stretching exponents of two rejuvenated samples
obtained by the two subsequent shear rate jumps:
$63\;\textrm{s}^{-1}\rightarrow223\;\textrm{s}^{-1}$ and
$223\;\textrm{s}^{-1}\rightarrow446\;\textrm{s}^{-1}$.
\begin{figure} \includegraphics[width=.40\textwidth]{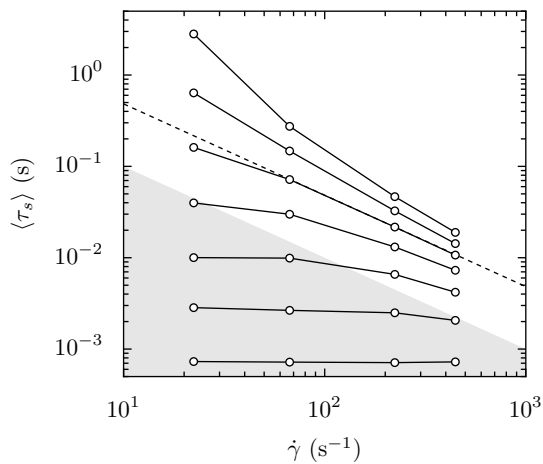} \caption{
Average slow relaxation time $\langle\tau_s\rangle$ as a function of shear rate
$\dot{\gamma}$ for seven evenly spaced waiting times between $0$ and $30$
hours.  Grey area represents the half-plane
$\langle\tau_s\rangle<\dot{\gamma}^{-1}$. Dashed line is a fit to a $\dot\gamma^{-1}$
power law.}\label{avtaugamma} \end{figure}

It is also instructive to look at the values of  $\langle\tau_s(\dot{\gamma},t_w)\rangle$
for constant $t_w$ as a function of $\dot{\gamma}$,  as reported in Fig. \ref{avtaugamma}.  
The grey region represents the half-plane $\langle\tau_s\rangle<\dot{\gamma}^{-1}$ where
slow dynamics takes place on a timescale shorter than $\dot{\gamma}^{-1}$.
This region is not affected by the presence of shear: the fluid is Newtonian and,
similarly to viscosity, $\langle\tau_s\rangle$ is not dependent on $\dot{\gamma}$.
On the other hand shear plays an important role in the complementary half-plane
where $\langle\tau_s\rangle$ displays a strong sensitivity to $\dot{\gamma}$.
The non-Newtonian behaviour in the upper half-plane of Fig. \ref{avtaugamma} 
resembles the same $\dot\gamma^{-1}$ power law (dashed line) observed in
rheological measurements of Laponite viscosity \cite{bonnreju}.

The whole scenario depicted above provides a microscopic counterpart of the
strong shear-thinning behaviour observed in rheological studies of Laponite
\cite{bonnreju} and of many other soft materials.
In both real \cite{larson} and simulated \cite{yamamoto, berthier} liquids,
the viscosity crossover from a Newtonian to a non-Newtonian 
regime (power law dependence on $\dot\gamma$), is usually found to be described
by scaling laws such as:
\begin{equation}\label{etascaling}
\eta(\dot{\gamma})\simeq\frac{\eta(0)}{1+\dot{\gamma}\tau_\eta} \end{equation}
Since viscosity is related to structural relaxation, one
could think, as suggested in \cite{yamamoto}, of a dynamical analogue of (\ref{etascaling}) in the form:
\begin{equation}\label{tauscaling}
\frac{1}{\tau(\dot{\gamma})}\simeq\frac{1}{\tau(0)}+A \dot{\gamma} \end{equation}
Which has also the advantage of an easily readable meaning: shear rate provides
a parallel relaxation channel to the system which becomes predominant as soon
as $1/\tau(0,T)\ll A\dot{\gamma}$. This simple relation is indeed found to work
remarkably well in computer models for supercooled liquids \cite{yamamoto}.
Using the same kind of reasoning, one would expect that even in an aging sample,
as soon as the unperturbed relaxation time grows large enough, shear rate will
fix the relevant timescale and the system will become stationary even in the
non ergodic phases.  This is actually what has been observed in a number of
recent numerical and theoretical papers, and also by rheological studies.  

We are now in the position of testing (\ref{tauscaling}) by directly investigating
the complete time behaviour of density dynamics in the presence of a shear flow.
At a first glance one would conclude that Fig. \ref{taubeta} is in contradiction
with (\ref{tauscaling}) and the expectation that shear stops aging. It is,
infact evident that relaxation time continues to grow, even if no longer
exponentially fast, at least for more than one order of magnitude since it
first ``feels" the shear field.  On the other hand, as shown in Fig. \ref{taubeta}, when the
crossover occurs,   the value of the stretching exponent is about $0.5$ or
less.  This means that
the unperturbed dynamics occurs on  a broad spectrum of timescales (spanning
two decades at least) and that $\langle\tau_s\rangle$ only represents an
average relaxation time.  Many approaches to slow dynamics in complex systems
\cite{hetero} suggest that such
a broad spectrum of time scale arises  from the heterogeneous character of slow
dynamics. Assuming this heterogeneous scenario, shear is only effective in
interrupting aging for those timescales growing longer than a fixed, shear
dependent, cutoff. In other words we expect that (\ref{tauscaling}) holds separately for every
timescale composing structural relaxation. The above scenario qualitatively
accounts for both the observations of a slowly growing (non saturating)
structural relaxation time after the crossover and a narrowing in the
distribution of timescales with respect to the unperturbed case.  We want to
push now the above observations on a more quantitative ground. Assuming that
(\ref{tauscaling}) holds separately for each timescale and calling $G(t_w, \tau)$
the unperturbed distribution of timescales \begin{equation}\label{G}
F_q^0(t_w,t)=\int_0^\infty G(t_w,\tau) \exp[-t/\tau]d\tau \end{equation} we would expect
that shear affects  (\ref{G}) as follows
\begin{eqnarray}\label{Gs} F_q^{\dot{\gamma}}(t_w,t)&=&\int_0^\infty G(t_w,\tau) \exp[-t(1/\tau+
A\dot{\gamma})]d\tau\\ &=&F_q^0(t_w,t)\exp[-t A\dot{\gamma}] \end{eqnarray}
In other words if we divide the correlation function measured  at a given
waiting time $t_w$ and shear rate $\dot{\gamma}$ by the corresponding
unperturbed ($\dot{\gamma}=0$) correlation measured at the same $t_w$ and
plot the result as a function of $\dot{\gamma}t$, we should obtain the master
curve $\exp[-A\dot{\gamma}t]$. 
\begin{figure}[t] \includegraphics[width=.40\textwidth]{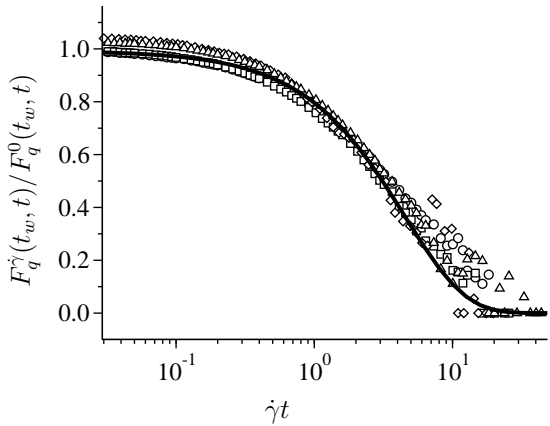} \caption{ Ratio
$F_q^{\dot{\gamma}}(t_w,t)/F_q^{0}(t_w,t)$ as a function of $\dot{\gamma}t$,
where $F_q^{\dot{\gamma}}(t_w,t)$ is the intermediate scattering function
measured after a waiting time $t_w$ during aging with applied shear rate
$\dot{\gamma}$.  $\circ$: $\dot{\gamma}$=446 s$^{-1}$, $t_w$=10 $h$ $\square$:
$\dot{\gamma}$=446 s$^{-1}$, $t_w$=14 $h$ $\vartriangle$: $\dot{\gamma}$=223
s$^{-1}$, $t_w$=14 $h$ $\diamond$: $\dot{\gamma}$=67 s$^{-1}$, $t_w$=14 $h$.
Solid line is an exponential fit $\exp[-A\dot{\gamma}t]$ with $A=0.22$.
}\label{ratio} \end{figure}

In Fig. \ref{ratio} we report the result of  such a procedure obtained for
three different values of $\dot{\gamma}$ and two waiting times. All curves
collapse on the same master curve which is well represented by a simple
exponential with $A=0.22$ (solid line).  Similarly, we could predict the shape
of relaxation for  given $t_w$ and $\dot{\gamma}$ by simply multiplying  the
unperturbed correlation function for the same $t_w$ by the function
$\exp[-A\dot{\gamma}t]$. If we do this for the $\dot{\gamma}$ values here
investigated and fit the result with a stretched exponential we obtain a
prediction for $\tau_s(t_w,\dot{\gamma})$ and $\beta(t_w,\dot{\gamma})$.
\begin{figure}[t] \includegraphics[width=.40\textwidth]{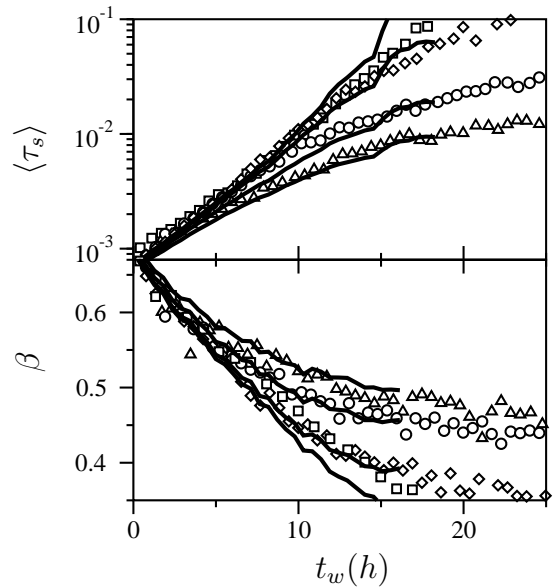} \caption{
Average relaxation time (top frame) and stretching parameter (bottom frame) as
a function of waiting time for different applied shear rates. Symbols are data
in Fig. \ref{taubeta}, solid line is the prediction of the model
described in the text.  }\label{taubetamodel} \end{figure}
The results of such a procedure, for those $t_w$ values where $F_q^0(t_w,t)$ is available,
are shown in Fig. \ref{taubetamodel} as solid lines.  
The overall agreement with the directly measured data points (open symbols) is very
satisfactory and supports the picture of the slow non-equilibrium dynamics of
Laponite as an heterogeneous superposition of relaxing units each independently
coupled to shear. 

In conclusion, we investigated the effect of shear on the non-equilibrium
structural dynamics of an aging  colloidal suspension of Laponite. The presence of a
shear flow strongly affects dynamics as soon as structural relaxation enters
the timescale $\dot{\gamma}^{-1}$.  In this shear dominated region the shear
rate dependence of the average slow relaxation time $\langle\tau_s\rangle$ is
well approximated by the power law $\dot{\gamma}^{-\alpha}$ with $\alpha\sim1$.
The effect of shear on the detailed shape of the intermediate scattering
function can be well described assuming that the slow relaxation process arises
from the heterogeneous superposition of many relaxing units each one
independently coupled to shear rate with a parallel composition rule for
timescales: $1/\tau\rightarrow1/\tau+A\dot\gamma$. 
The authors wish to thank F.~Zamponi for helpful discussions.


\end{document}